\documentstyle[11pt,newpasp,twoside,epsf]{article}
\reversemarginpar

\begin{document}
\title{Comparative CMBology: Putting Things Together}
\author{Charles H. Lineweaver}
\affil{School of Physics, University of New South Wales, Sydney, Australia\\
 charley@bat.phys.unsw.edu.au}
\begin{abstract}
I present a series of diagrams which illustrate why the cosmic microwave background (CMB) data 
favor certain values for the cosmological parameters.
Various methods to extract these parameters from CMB and non-CMB observations 
are forming an ever-tightening network of interlocking constraints.
I review the increasingly precise constraints in the $\Omega_{m} - \Omega_{\Lambda}$ plane and 
show why more cosmologists now prefer $\Lambda$CDM cosmologies to any other leading model.
\end{abstract}
\suppressfloats
\vspace{-7mm}
\section{What is the CMB data trying to tell us?}
\begin{figure}[!ht]
\plotfiddle{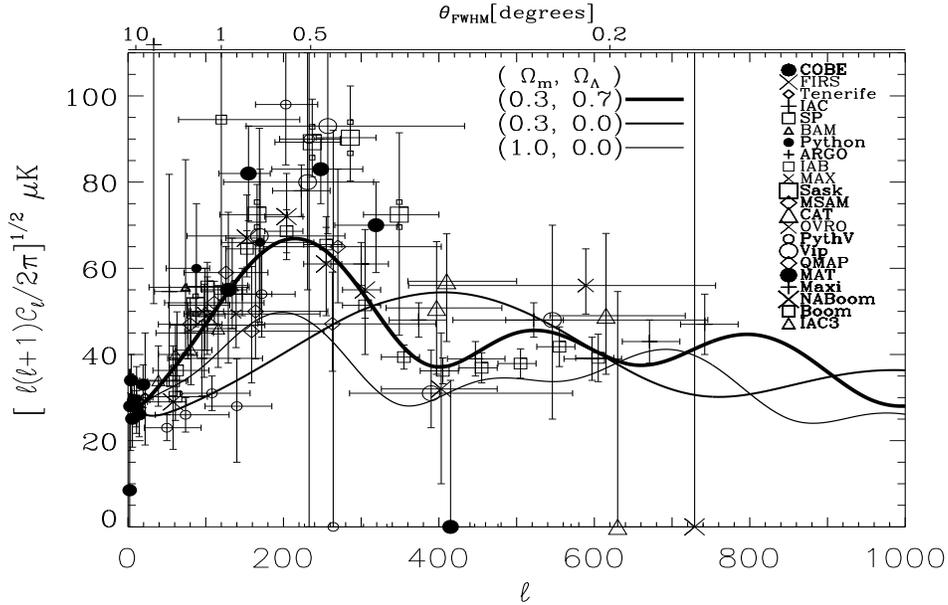}{7.0cm}{0}{63}{53}{-187}{-57}
\caption{{\footnotesize Compilation of recent CMB observations from 22 experiments compared to three popular
cosmological models. At the last IAU, all three models were serious contenders for reality.
This is no longer the case.}}
\label{fig:data}
\end{figure}

\begin{figure}[!hb]
\plotfiddle{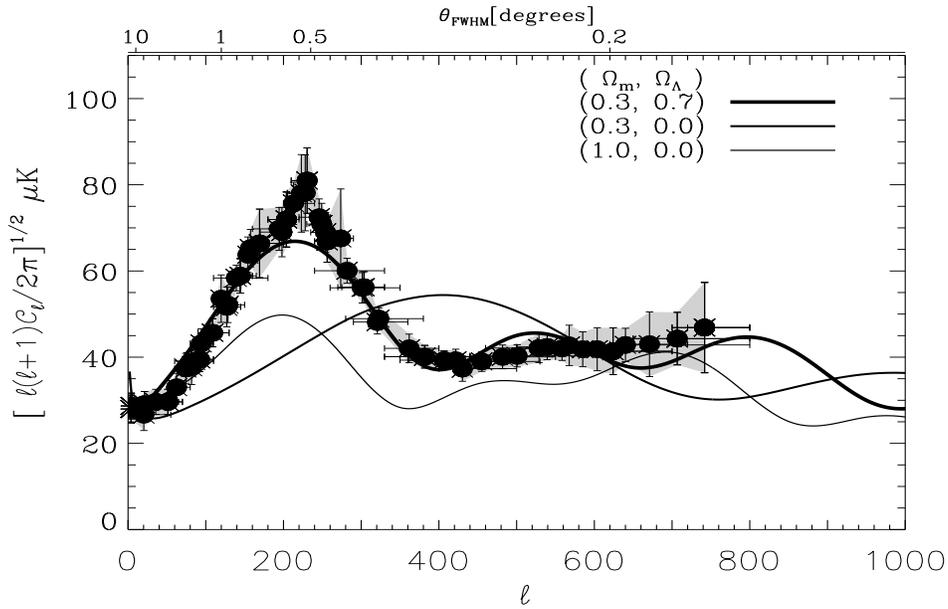}{7.0cm}{0}{63}{53}{-187}{-57}
\caption{{\footnotesize Same as Fig.~\ref{fig:data}, but the data points have been binned and averaged and rebinned multiple times
to provide a series of points and a grey area defined by the 
1 $\sigma$ error bars of the binned points.
The points are not independent.
The well defined position of the first peak in the spectrum at $\ell_{p} \sim 220$ easily rules out the 
$(\Omega_{m}, \Omega_{\Lambda}) = (0.3,0.0)$ model shown which peaks at $\ell_{p} \sim 400$
(see e.g. Lineweaver 1998).}}
\label{fig:binnedata}
\end{figure}

\begin{figure}[!ht]
\plotfiddle{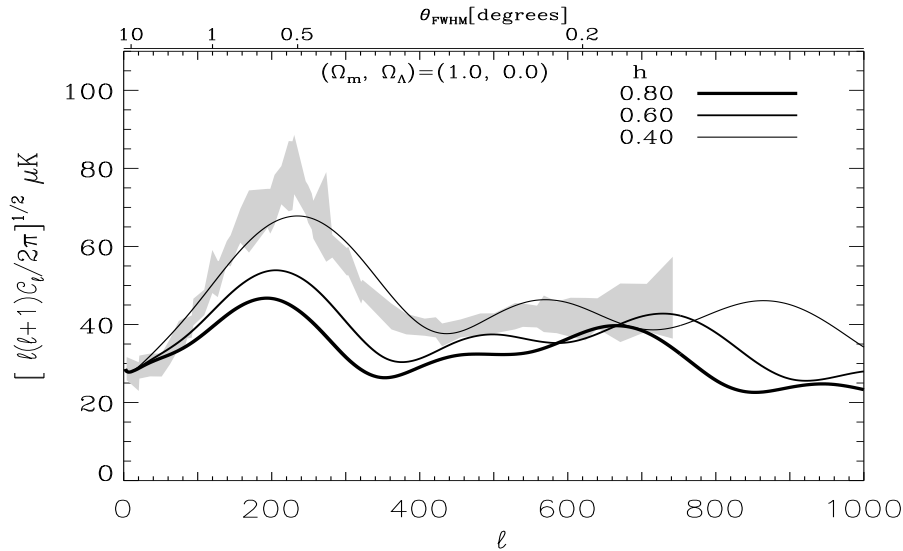}{6.3cm}{0}{60}{48}{-175}{-52}
\caption{{\footnotesize {\bf Standard CDM is consistent with CMB data, but only if $h$ is extremely low.}
Until recently, the standard cold dark matter model ($\Omega_{m}, \Omega_{\Lambda}) = (1,0)$
was the leading cosmological candidate, but the amplitude of the first peak is too low in these
models unless $h \sim 0.40$.
See the $h$ contours in Panel A of Figure~\ref{fig:science} for more details.
The grey region is preferred by the data and comes from the previous figure.}}
\label{fig:highh}
\end{figure}

\begin{figure}[!hb]
\plotfiddle{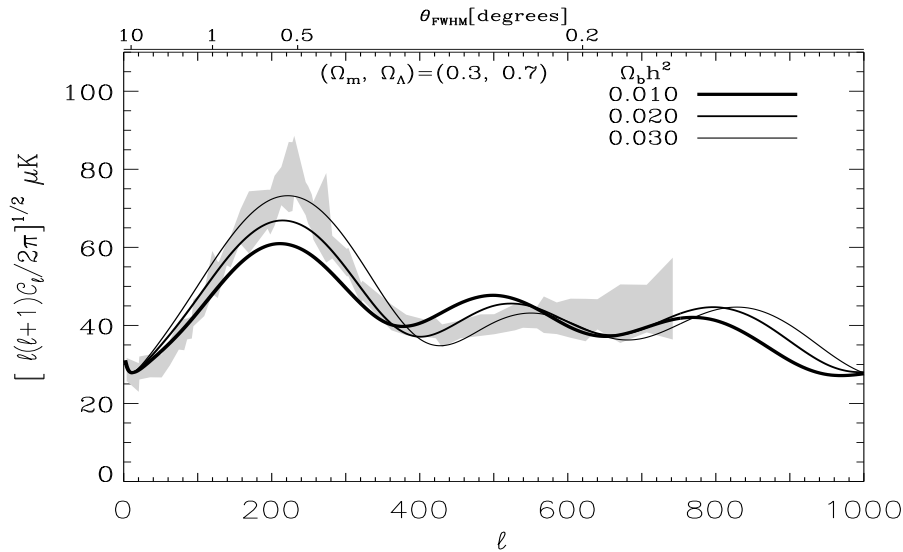}{6.3cm}{0}{60}{48}{-175}{-52}   
\caption{{\footnotesize {\bf Why does the CMB prefer high values of $\Omega_{b} h^{2}$?} 
Compared to big-bang-nucleosynthesis-dependent estimates ($\Omega_{b} h^{2} \approx 0.020$ 
e.g. Tytler et al. 2001)
the first peak of the CMB data has a slightly higher amplitude and the second peak has a slightly lower amplitude.
Raising $\Omega_{b} h^{2}$ to 0.030 fits the data better because it raises the first peak and lowers the second peak.
More precise data in the $\ell \sim 500$ region will soon be available to help solve this tentative conflict.
}}
\label{fig:omegabaryon}
\end{figure}

\begin{figure}[!ht]           
\plotfiddle{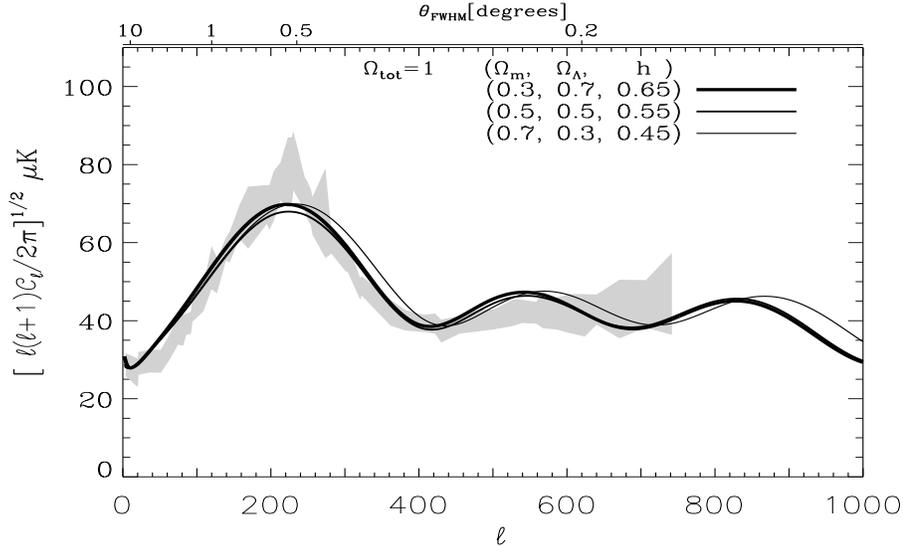}{6.5cm}{0}{60}{48}{-187}{-57}       
\caption{{\footnotesize {\bf Degeneracy is a big problem.}
These three flat models ($\Omega_{tot} = \Omega_{m} + \Omega_{\Lambda} = 1, \Omega_{k} = 0$) have 
very different values for $\Omega_{m}$ and $\Omega_{\Lambda}$,
but the CMB cannot differentiate between them if $h$ can vary .
Reducing $h$ raises the peak (Fig.~\ref{fig:highh}) while increasing $\Omega_{m}$ lowers the peak.
Similar but usually more subtle degeneracies become more numerous as the dimensionality of explored parameter space increases.
An important issue is what range of values does one allow for $h$ and other parameters.}}
\label{fig:degeneracy}
\end{figure}

\begin{figure}[!hb]
\plotfiddle{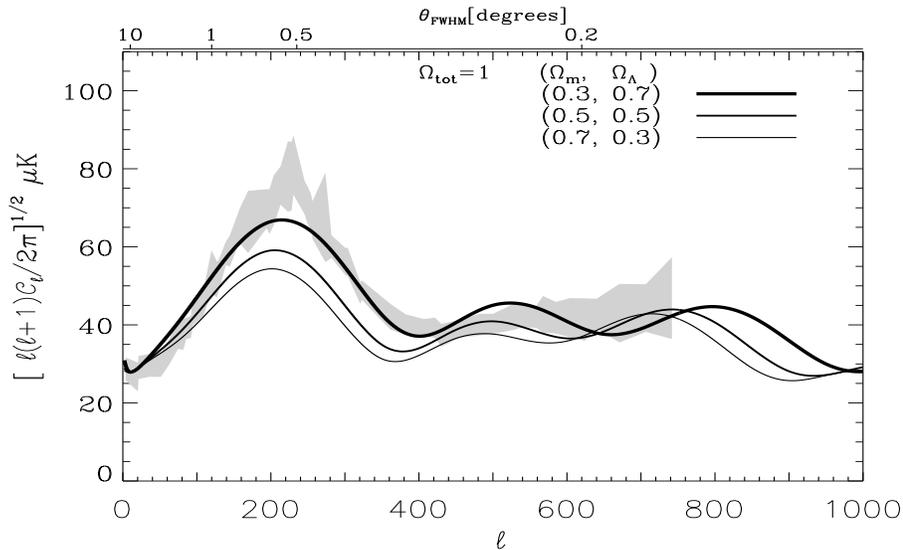}{6.5cm}{0}{60}{48}{-187}{-57}      
\caption{{\footnotesize {\bf Degeneracies can be broken.} Precise non-CMB constraints (such as an independent measurement of $h$) 
or much more precise CMB data can break degeneracies.
Here, setting $h = 0.70$ breaks the degeneracy of Fig. 5, allowing the CMB 
to discriminate between various flat models.}}
\label{fig:breakdegeneracy}
\end{figure}

\begin{figure}[!ht]
\plotfiddle{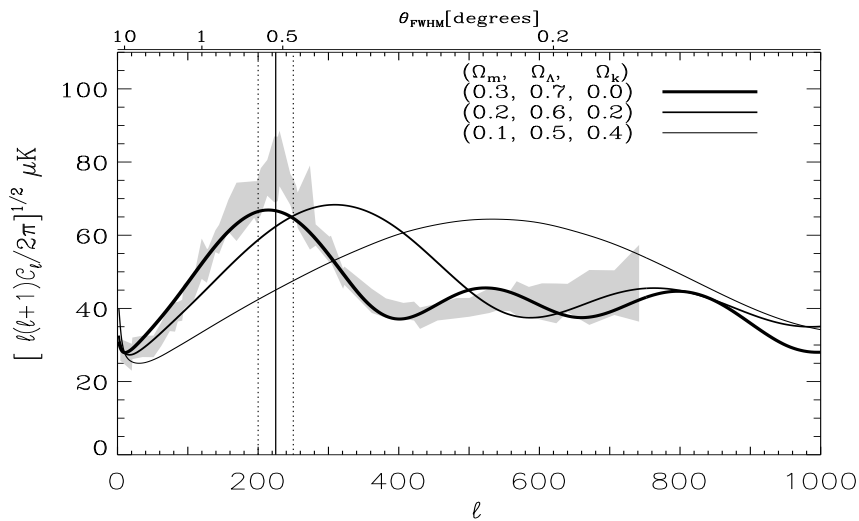}{5.8cm}{0}{57}{45}{-175}{-52}
\caption{{\footnotesize {\bf The CMB can measure the geometry of the Universe, $\Omega_{k}$,
better than other cosmological observations.}
The position of the first peak in the CMB data, $\ell_{p}$, is probably the
 best measurement we have of the geometry of the universe.
It is an excellent tool to answer the question: `Is the universe spatially open, flat or closed?'
Or, in terms of cosmological parameters ($\Omega_{k} = 1- \Omega_{tot}$), `Is $\Omega_{k}$ greater than, equal to or less than zero?'
(or equivalently, `Is $\Omega_{tot}$ less than, equal to or greater than one?')
We have set $\Omega_{k} = 0.0, 0.2, 0.4$ (flat, open, more open, respectively). 
The vertical lines indicate the peak in the data (Fig.~\ref{fig:zoomin}).}}
\label{fig:geometry}
\end{figure}

\begin{figure}[!hb]
\plotfiddle{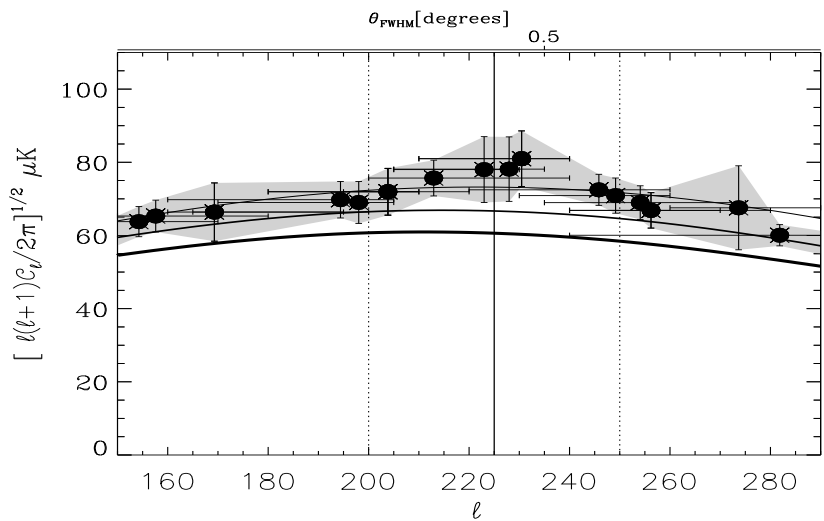}{5.8cm}{0}{57}{45}{-175}{-52}
\caption{{\footnotesize {\bf Where is the peak?} 
This is a blow-up of the region of the first peak with the multiply-binned
points from Fig.~\ref{fig:binnedata} and the models from Fig.~\ref{fig:omegabaryon}.
An eye-ball model-independent estimate of the position of the first peak is $\ell_{p} \approx 225 \pm 25$, 
indicated by the vertical lines (same as in previous figure). 
This result is based on the data shown in Fig.~\ref{fig:data}. The points are {\it not} independent.}}
\label{fig:zoomin}
\end{figure}

A convenient way to interpret CMB observations is to fit the angular power spectrum of the data to 
parameter-dependent models. Figure \ref{fig:data} shows the recent CMB measurements along with three such
models. In Figure~\ref{fig:binnedata}, binning of this data reduces the scatter and provides a representative region
favored by the data.
Important parameters that can be constrained by CMB power spectra include Hubble's constant $h$,
the cosmological constant $\Omega_{\Lambda}$, the density of cold dark matter $\Omega_{\rm CDM}$, and
the density of baryonic matter $\Omega_{b}$.
Figures~\ref{fig:data}~-~\ref{fig:zoomin} provide a qualitative feel for the lever arm that the CMB data provides 
for constraining these and other  parameters simultaneously.
Unless stated otherwise, the models shown have the following default values:
$h= 0.70$, $\Omega_{\Lambda}=0.7$, $\Omega_{m}= \Omega_{\rm CDM} + \Omega_{b} =0.3$,  
$\Omega_{b}h^{2} = 0.020$, a power spectral index of primordial scalar density fluctuations $n_{s}=1$ 
and an overall normalization $Q_{10} = 18\;\mu$K.
The grey band in Figure~\ref{fig:binnedata} is reproduced in Figures~\ref{fig:highh}~-~\ref{fig:zoomin} 
and represents the data in a model-independent way. With it, the eye can pick out which models best fit the data. 

A reduction in $h$ increases the amplitude of the first peak (Fig.~\ref{fig:highh}).
An increase in the number of baryons increases the gravitating mass of the oscillating 
baryon-photon fluid .
This enhances the first peak (Fig.~\ref{fig:omegabaryon}) by producing more gravitational compression as the baryons drag 
the photons further into the potential wells  (and further away from the potential hills). 
The second peak is suppressed because, before decoupling, these smaller scales experienced the same additional 
compression (and rarefaction) and, at decoupling, we are seeing a subdued rebound from this enhanced 
compression (and rarefaction), i.e., we are seeing the smaller amplitude of an oscillation whose zero level had 
been lowered in the previous oscillation by the effect of additional baryons.
An increase in $\Omega_{m}$ decreases the amplitude of the first peak (Fig.~\ref{fig:breakdegeneracy}). 
For discussion of the physics of the parameter dependencies of features in the CMB power spectrum  
see e.g. Hu \& Sugiyama (1995), Hu (1995), Tegmark (1996), Lineweaver et al. (1997).

\clearpage

\clearpage

\clearpage

\clearpage

\clearpage

Just as the SNIa data is our strongest lever arm for determining the acceleration of the universe, $q_{o}$,
the CMB data is our strongest lever arm for determining the geometry of the universe, $\Omega_{k}$.
The ability of the CMB to constrain $\Omega_{k}$ can be seen in Fig.~\ref{fig:science}, panel A and Fig.~\ref{fig:latest} 
in which the CMB contours are elongated in the $\Omega_{k} = $ {\it constant} direction (upper left to lower right).
However, since both $\Omega_{k}$ and $h$ 
control the position of the peak, 
there is a slight degeneracy. It is difficult to separate the effect of the spatial geometry of the universe from 
the effect of $h$.
This degeneracy is reflected in the width of the elongated CMB constraints in 
the $\Omega_{m} - \Omega_{\Lambda}$ plane. The models in Fig.~\ref{fig:zoomin} have $\Omega_{k} = 0$, $h = 0.70$. 
The data and the high baryon model peak at $\ell \approx 225$. 
This crude eye-ball estimate should be compared to the more careful but model-dependent estimates of
Bond et al. (2001, $\ell_{p} = 212 \pm 7$) 
and Hu et al. (2001, $\ell_{p} < 218$ at the $2\sigma$ level based on the Boomerang and Maxima results only).

\section{Putting It All Together}

\begin{figure}[!ht]
\plotfiddle{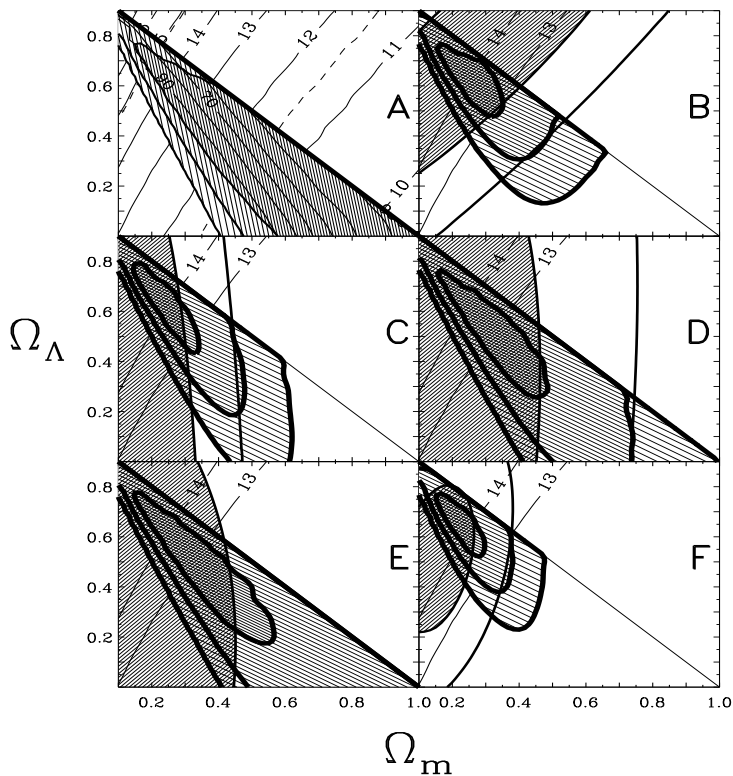}{9.2cm}{0}{60}{50}{-180}{-20}
\caption{{\footnotesize
The regions of the $(\Omega_{m}, \Omega_{\Lambda})$ plane preferred by
various constraints. 
({\bf \sf A}) CMB, ({\bf \sf B}) SNe, 
({\bf \sf C}) cluster mass-to-light ratios, ({\bf \sf  D}) cluster abundance evolution, 
({\bf \sf E}) double radio lobes,  and ({\bf \sf F}) all combined (see Lineweaver 1999 for details).
The power of combining the CMB constraints from {\bf \sf A} with each of the other constraints is also shown.
In {\bf \sf  A}, the elongated triangles (from upper left to lower right)  are the approximate 
1$\sigma$, 2$\sigma$ and 3$\sigma$ confidence levels of the likelihood from CMB data.
{\bf {\sf A}} also shows the $h$ contours which maximize the likelihood for the given values of $\Omega_{m},\Omega_{\Lambda}$.
Notice the best-fiting $h$ values in panel A for models close to $(\Omega_{m}, \Omega_{\Lambda}) = (1,0)$
are less than $0.40$ (see Fig.~\ref{fig:highh}).
The thick contours in {\bf{\sf  F}} show the region preferred by the combination of the separate constraints shown in ({\bf \sf A}) 
through ({\bf \sf E}).
In ({\bf{\sf  A}}), the thin iso-$t$ contours (labeled ``10'' through ``14'') indicate the age in Gyr  when $h = 0.68$ 
is assumed. For reference, the 13- and 14-Gyr contours are in all the panels.
}}
\label{fig:science}
\end{figure}

Presumably we live in a universe which corresponds to a single point in multidimensional parameter space.
If the universe is to make sense, independent determinations of $\Omega_{m}$, $\Omega_{\Lambda}$, $h$ and 
the minimum age of the Universe must be consistent with each other (Fig.~\ref{fig:ages}).
Estimates of $h$ from HST Cepheids and from the CMB must overlap.
Deuterium and CMB determinations of $\Omega_{b} h^{2}$ should be consistent (Fig.~\ref{fig:omegabaryon}, 
but see Kaplinghat \& Turner 2000).
%
Regions of the $\Omega_{m} - \Omega_{\Lambda}$ plane favored by SNIa and CMB must overlap with each other and with
other independent constraints. That this is the case is shown in Fig.~\ref{fig:science}.


\subsection{Consistency enforcement: a worked example in the $\Omega_{m} - \Omega_{\Lambda}$ plane}

\begin{figure}[!ht]
\plotfiddle{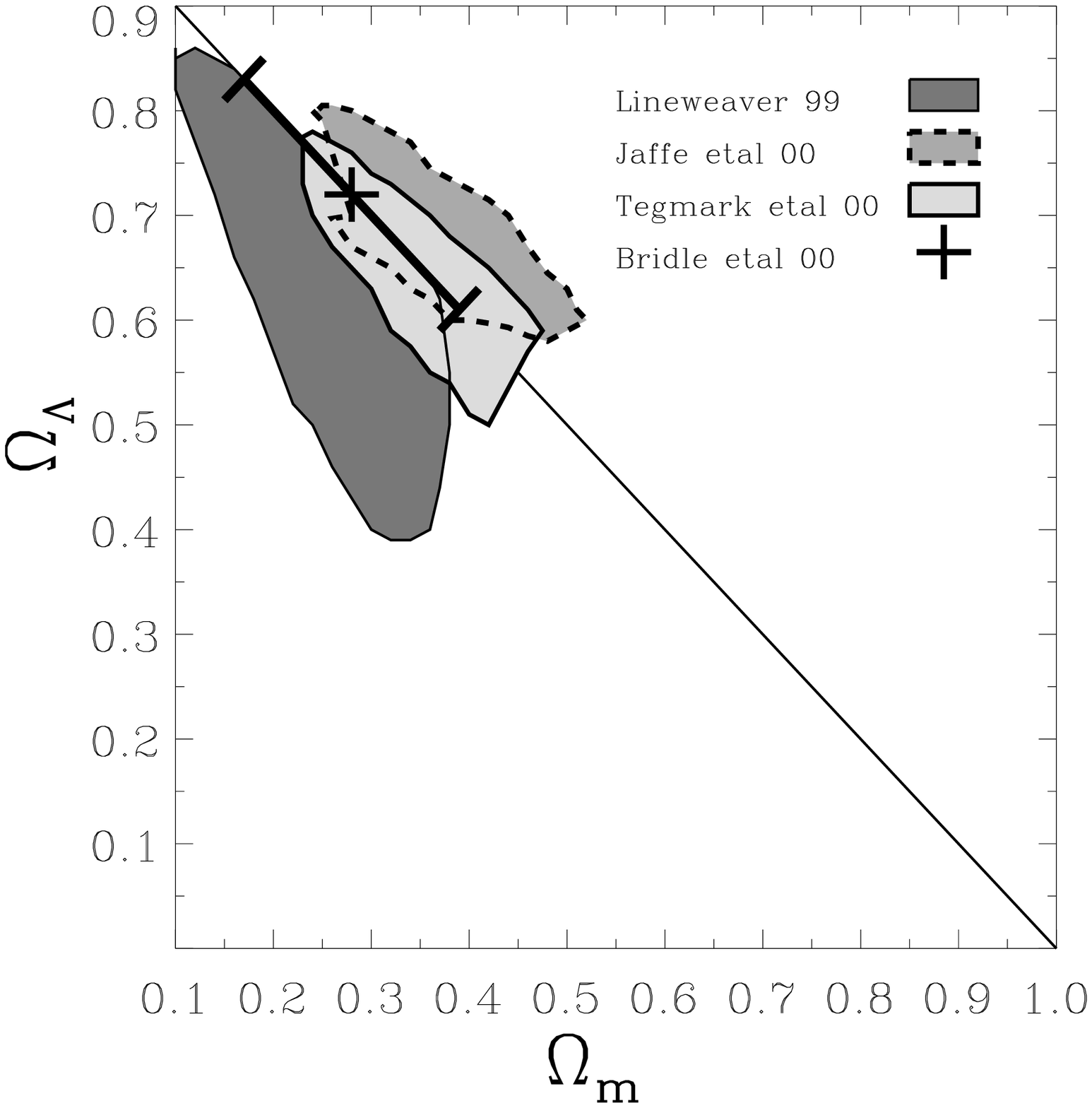}{9.5cm}{0}{67}{57}{-210}{-70}
\caption{{\footnotesize {\bf Joint Constraints in the $\Omega_{m} - \Omega_{\Lambda}$ plane.}
The reported $2\sigma$ constraints from Lineweaver (1999), Jaffe et al. (2000), Bridle et al. (2000) and 
Tegmark et al. (2000) are shown. Each of these papers is based on CMB data but used different methods and 
different combinations of non-CMB constraints.
A failure to reach a disagreement is evident. 
}}
\label{fig:latest}
\end{figure}

\begin{figure}[!ht]
\plotfiddle{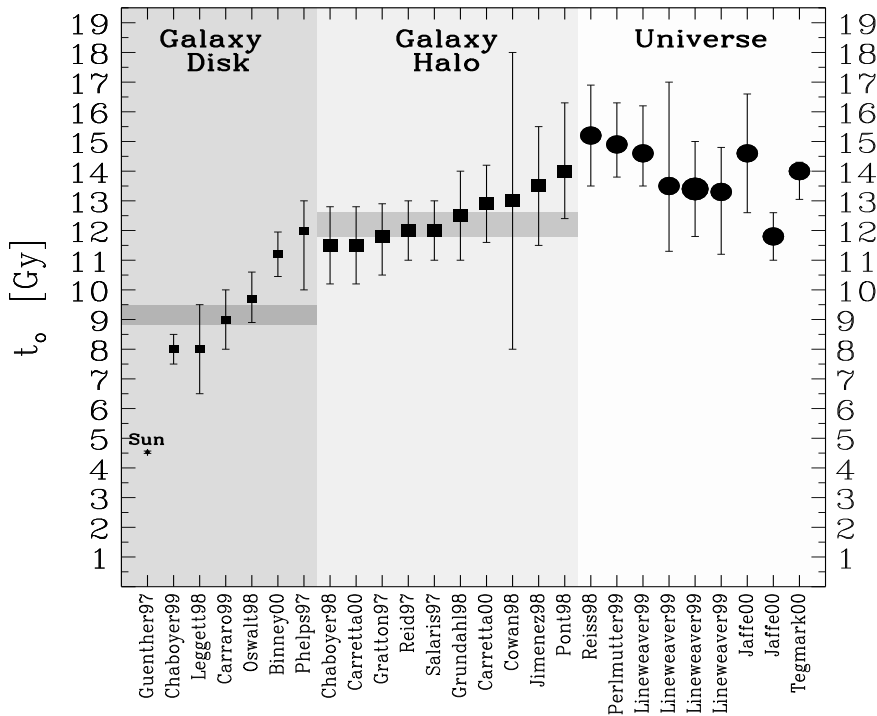}{9.3cm}{0}{65}{55}{-190}{-35}
\caption{{\footnotesize {\bf The Universe is finally older than our Galaxy!}
Age estimates of the Universe based on the new estimates of $h$, $\Omega_{m}$ and $\Omega_{\Lambda}$
are larger than the age estimates of the oldest objects in our galaxy. This has not always been the case and has been
the source of some embarassment. Age results from Binney et al. (2000), Carretta et al. (2000), Jaffe et al. (2000) and 
Tegmark et al. (2000) have been added to this figure taken from Lineweaver (1999).}}
\label{fig:ages}
\end{figure}

If any $\Omega_{\Lambda} = 0$ model can squeak by the SNIa constraints it is the very low $\Omega_{m}$ models.
However these models are the ones most strongly excluded by the CMB data (Fig.~\ref{fig:science}, panels A \& B).
The SNIa results show that the universe is accelerating but cannot yield a value of the cosmological constant 
unless one assumes that the universe is flat. However, that assumption
is not necessary since we have CMB data that tells us that the universe {\it is} approximately flat.
Figure~\ref{fig:science} is an example of how various independent constraints can be combined with
CMB constraints in the $\Omega_{m} - \Omega_{\Lambda}$ plane. The 2 $\sigma$ combined constraints in panel
F are limited to open and flat models and are reproduced in Fig.~\ref{fig:latest}.

All four constraints in Fig.~\ref{fig:latest} come from CMB constraints
assuming adiabatic initial conditions, in combination with SNIa constraints.
The amplitude ($\sigma_{8}\Omega_{m}^{0.6}$) and shape ($\Gamma$) of the local 
power spectrum of galaxies were included as additional constraints
by Bridle et al. (2000, catalogue of peculiar velocities), Tegmark et al. (2000, IRAS PCSz)
and Jaffe et al. (2000, $\sigma_{8}\Omega_{m}^{.56} = 0.55^{+0.02}_{-0.02}{}^{+0.11}_{-0.08}$
$\Gamma + (n_{s} -1)/2 = 0.22^{+0.07}_{-0.04}{}^{+0.08}_{-0.07}$
where the first errorbar is the Gaussian prior and the second is the full range considered).
To constrain $\Omega_{m}$ and $\Omega_{\Lambda}$, marginalization of the other parameters was done by
maximization (Lineweaver 1999, Tegmark et al. 2000) and integration (Bridle et al. 2000, Jaffe et al. 2000).
The number of parameters marginalized over to obtain the results shown in Fig.~\ref{fig:latest} is given in Table
1 along with the priors used for $h$, $\Omega_{b}h^{2}$ and geometry.
%
Calibration errors were treated slightly differently.


{\footnotesize
\begin{table}[!hb]
\vspace{-7mm}
\begin{center}
\caption{{\footnotesize Data and conditions used in four papers deriving the four constraints in the 
$\Omega_{m} - \Omega_{\Lambda}$ plane shown in Fig.~\ref{fig:latest}.}}
\begin{tabular}{l c c c c l} 
\multicolumn{1}{l}{Reference} &
\multicolumn{1}{c}{Data} &
\multicolumn{1}{c}{\# parameters} &
\multicolumn{1}{c}{$h$} &
\multicolumn{1}{c}{$\Omega_{b}h^{2}$} &
\multicolumn{1}{c}{Geometry}\\
\hline
Lineweaver(1999)    & pre 4/1999      &       6        &$ 0.68 \pm 0.10   $&$0.015 \pm 0.01  $& flat open        \\
Jaffe et al.(2000)  & DMR+Boom+Max    &       7        &$ [0.45,\;0.90]   $&$[0.003,\;0.02]  $& flat open closed \\
Bridle et al.(2000) & all             &       3        &$ [0.30,\;0.90]   $&$0.019 \pm 0.00  $& flat             \\
Tegmark et al.(2000)& all             &      10        &$ 0.74 \pm 0.08   $&$0.020 \pm 0.00  $& flat open closed \\
\end{tabular}\\
\end{center}
\end{table}
}
\normalsize

\section{The New Standard Model}
The parameters that we are trying to determine are fundamental because they give the answers to 
fundamental questions: What is the universe made of? How old is the universe? How big is the observable universe?
My view of the best current values are given in Table 2.

{\footnotesize
\begin{table}[!h]
\vspace{-5mm}
\begin{center}
\caption{{\footnotesize Fundamental Parameters}}
\begin{tabular}{l l} 
\multicolumn{1}{l}{Parameter estimate} &
\multicolumn{1}{l}{References}\\
\hline
$\Omega_{m}        = 0.3   \pm   0.1      $&    Fig.~\ref{fig:latest}     \\
$\Omega_{\Lambda}  = 0.7   \pm   0.1      $&    Fig.~\ref{fig:latest}     \\
$h                 = 0.70  \pm   0.1      $&    Mould et al 99, Parodi et al 00, Freedman et al 01\\
$t_{o}             = 13.4  \pm 1.6$  Gyr   &    Fig.~\ref{fig:ages}      \\
$\Omega_{b} h^{2}  = 0.020  \pm   0.010    $&   Olive et al 99,  Tytler et al 01\\
\end{tabular}\\
\end{center}
\end{table}
}    

In this paper entitled `putting things together' I have left out a lot: tensor modes ($A_{t},n_{t}$), early reionization ($\tau$),
hot dark matter ($\Omega_{\nu}$, do neutrinos have a cosmologically significant mass), non-Gaussianity, 
topological defects, quintessence,  scale dependent slopes ($n_{s}(\ell)$), non-adiabatic 
initial conditions, variation in the speed of light and/or the fine structure constant. 
Any or all of these, or some we have not thought of, may prove to be crucial in the high precision future of CMBology.

\subsection{$\Lambda$CDM! Any objections?}
Lensing constraints in the $\Omega_{m} - \Omega_{\Lambda}$ plane have been invoked as evidence against
$\Lambda$CDM models (e.g. Kochanek 1996). However in his contribution to these proceedings based on the JVAS/CLASS survey,
Helbig reports new lensing constraints which do not exclude the $(\Omega_{m}, \Omega_{\Lambda}) = (0.3, 0.7)$ model.

Numerical simulations of the central density profiles of low surface brightness galaxies in $\Lambda$CDM models 
do not match the observations very well,  but this may be the result of our ignorance of how baryons assemble
into galaxies (Navarro \& Steinmetz 2001).

Some cosmologists believe there may be a coincidence problem. Why just now should we have $\Omega_{\Lambda} \sim \Omega_{m}$?
(e.g. Carroll 2001, Fig. 11). However, this `coincidence'  may well be explained away as an anthropic selection effect.

Another problem with $\Lambda$CDM is that it is an observational result that has yet to be theoretically confirmed.
From a quantum field theoretic point of view, $\Omega_{\Lambda} \sim 0.7$ presents a huge problem and even `seems ridiculous' 
(Carroll 2001, Sect. 1.3, see also Cohn 1998 and Sahni \& Starobinsky 1999).
But if observations continue to yield $\Omega_{\Lambda} \approx 0.7$ some imaginative theorist will solve this problem.

As more cosmological data comes in, the CMB and non-CMB constraints form an ever-tightening network of interlocking constraints.
Fig.~\ref{fig:science} shows some of the pieces of this ever-tightening network while Fig.~\ref{fig:latest} shows the latest
tightenings.
One of the major new unsung results of recent efforts to 
simultaneously fit parameters is that the fits are good fits.
The parameter values are consistent with each other.  This may not last.
%
%
If, as our data gets better, the best fit
model is no longer a good fit, new ideas would be needed.
Parameter space may need another few dimensions to contain the real universe.
I know of no better way to find
these new dimensions than to analyse the increasing precise measurements of the CMB and combine the results with
other independent cosmological observations.
New data from Boomerang, Maxima, CBI, DASI, VSA, MAP, Beast, Planck and others ensure the health and guarantee
continued progress towards determining the fundamental parameters of our universe.


\vspace{-2mm}
\small
\thebibliography
\bibitem[Binney et al 2000]{Bin00} Binney, J., Dehnen, W., \& Bertelli, G. MNRAS, submitted, 2001, astro-ph/0003479 \vspace{-2mm} 
\bibitem[Bond et al 2001]{B01} Bond, J.R. et al. 2001, these proceedings,  astro-ph/0011378 \vspace{-2mm} 
\bibitem[Bridle et al. 2000]{Bridle00}Bridle, S.L., Zehavi, I., Dekel, A, Lahav, O., Hobson, M.P., Lasenby,A.N. 2000, MNRAS
astro-ph/0006170 \vspace{-2mm} 
\bibitem[Carretta et al. 2000]{Car00} Carretta, et al. 2000, \apj, 533, 215\vspace{-2mm} 
\bibitem[Carroll 2000]{C01} Carroll, S.M. 2001, Living Reviews of Relativity, submitted,  Fig. 11, astro-ph/0004075\vspace{-2mm} 
\bibitem[Cohn 1998]{C98} Cohn, J.D. 1998, Astrophys. Sp. Sci 259, 213, astro-ph/9807128\vspace{-2mm} 
\bibitem[Efstathiou et al. 1999]{E99}Efstathiou, G., Bridle, S.R., Lasenby, A.N., Hobson, M.P., Ellis, R.S. 1999, MNRAS, 303, L47-L52 \vspace{-2mm}  
\bibitem[Freedman et al 2001]{Fre01}Freedman, W. et al. 2001, in preparation (these proceedings?) \vspace{-2mm} 
\bibitem[Helbig 2001]{Hel01}Helbig, P. 2001, these proceedings, astro-ph/0011031 \vspace{-2mm} 
\bibitem[Hu 1995]{Hu95}Hu, W.  1995, PhD thesis, Berkeley\vspace{-2mm}
\bibitem[Hu \& Sugiyama 1995]{HS95}Hu, W. \& Sugiyama, N. 1995, \apj, 444, 489 \vspace{-2mm}
\bibitem[Hu et al. 2000]{Hu00}Hu, W., Fukugita, M., Tegmark, M. 2001, \apj, submitted \vspace{-2mm} 
\bibitem[Jaffe etal. 2000]{Jaffe00}Jaffe, A. et al. 2000, PRL, in press, astro-ph/0007333 \vspace{-2mm} 
\bibitem[Kaplinghat \& Turner 2000]{Kap00} Kaplinghat, M. \& Turner, M.S. submitted, astro-ph/0007452 \vspace{-2mm} 
\bibitem[Kochanek 1996]{K96} Kochanek, C. 1996, \apj, 466, 638\vspace{-2mm} 
\bibitem[Lineweaver etal 1997]{L97}Lineweaver, C.H. et al.  1997, \aap, 322, 365, Section 3.2.1 \vspace{-2mm} 
\bibitem[Lineweaver \& Barbosa 1998]{LB98}Lineweaver, C.H. \& Barbosa, D. 1998, \apj, 496, 624 \vspace{-2mm} 
\bibitem[Lineweaver 1998b]{L98b}Lineweaver, C.H. 1998, \apj, 505, L69 \vspace{-2mm} 
\bibitem[Lineweaver 1999]{L99} Lineweaver, C.H. 1999, Science, 284, 1503-1507 \vspace{-2mm} 
\bibitem[Mould 1999]{M99}Mould, J. et al. 1999, astro-ph/9909260 
\bibitem[Navarro \& Steinmetz 2001]{NS01}Navarro, J. F. \& Steinmetz, M. 2001, \apj, submitted, astro-ph/0001003 \vspace{-2mm} 
\bibitem[Olive 1999]{Oli99}Olive, K.A., Steigman, G. \& Walker, T.P. 1999, astro-ph/9905320 \vspace{-2mm} 
\bibitem[Parodi 2000]{Pa00}Parodi, B.R., Saha, A., Sandage, A. \& Tammann, G.A. 2000, \apj, submitted, astro-ph/0004063 \vspace{-2mm} 
\bibitem[Sahni 1999]{SS99} Sahni, V. \& Starobinsky, A. 1999, astro-ph/9904398 \vspace{-2mm} 
\bibitem[Tegmark 1996]{Teg96} Tegmark, M. 1996, in Proc. Enrico Fermi, Course CXXXII, Varenna, 1995 \vspace{-2mm} 
\bibitem[Tegmark, Zaldarriagga \& Hamilton 2000]{TZH00} Tegmark, M., Zaldarriaga, M. \& Hamilton, A.J.S. 2001, Phys. Rev. D.,
astro-ph/0008167 v3 \vspace{-2mm} 
\bibitem[Tytler 2001]{Tyt01}Tytler, D. et al. 2001, Physica Scripta, in press astro-ph/0001318 \vspace{-2mm}  
\endthebibliography

\end{document}